\pdfoutput=1

\documentclass[11pt]{article}
\usepackage{longtable}
\usepackage[preprint]{acl}
\usepackage[ruled,vlined]{algorithm2e}
\usepackage{times}
\usepackage{latexsym}
\usepackage{xcolor}

\usepackage{amsmath, amssymb, graphicx, multirow, multicol, comment, subcaption, url, float, etoolbox, adjustbox, pgf, soul, geometry, colortbl, booktabs, verbatim}
\usepackage[T1]{fontenc}
\usepackage[utf8]{inputenc}
\usepackage{microtype}
\usepackage{inconsolata}

\usepackage{microtype}
\definecolor{lightred}{rgb}{1.0, 0.8, 0.8}
\definecolor{lightblue}{rgb}{0.8, 0.9, 1.0}
\definecolor{lightgreen}{rgb}{0.8, 1.0, 0.8}
\definecolor{lightyellow}{rgb}{1.0, 1.0, 0.8}
\definecolor{lightpurple}{rgb}{0.9, 0.8, 1.0}
\definecolor{lightorange}{rgb}{1.0, 0.9, 0.8}

\title{SeQuiFi: Mitigating Catastrophic Forgetting in Speech Emotion Recognition with Sequential Class-Finetuning}

\author{
  Sarthak Jain$^{1*}$,
  Orchid Chetia Phukan$^{1}$\thanks{\footnotesize{Authors contributed equally as first authors}},
  Swarup Ranjan Behera$^{2*}$\\
  \textbf{Arun Balaji Buduru}$^{1}$,
  \textbf{Rajesh Sharma}$^{1,3}$\\
  \textsuperscript{1}IIIT-Delhi, India, 
  \textsuperscript{2}Reliance Jio AICoE , India,
  \textsuperscript{3}University of Tartu, Estonia\\ 
  \texttt{\textbf{Correspondence:} \textcolor{blue}{orchidp@iiitd.ac.in}}
}

\begin{document}
\maketitle

\begin{abstract}
In this work, we introduce \textbf{SeQuiFi}, a novel approach for mitigating catastrophic forgetting (CF) in speech emotion recognition (SER). \textbf{SeQuiFi} adopts a sequential class-finetuning strategy, where the model is fine-tuned incrementally on one emotion class at a time, preserving and enhancing retention for each class. While various state-of-the-art (SOTA) methods, such as regularization-based, memory-based, and weight-averaging techniques, have been proposed to address CF, it still remains a challenge, particularly with diverse and multilingual datasets. Through extensive experiments, we demonstrate that \textbf{SeQuiFi} significantly outperforms both vanilla fine-tuning and SOTA continual learning techniques in terms of accuracy and F1 scores on multiple benchmark SER datasets, including CREMA-D, RAVDESS, Emo-DB, MESD, and SHEMO, covering different languages.

\end{abstract}

\section{Introduction}
Speech Emotion Recognition (SER) is a task that has far-reaching implications, from enhancing human-computer interactions to improving diagnostic tools in healthcare settings. Despite notable advancements in SER methodologies, a significant challenge persists, the generalization of models across diverse data distributions and languages.
Variations in acoustic properties, linguistic structures, and emotional expressions across different speakers and cultures lead to substantial performance degradation when models trained on a specific dataset or language are deployed in another context. \par

A commonly preferred strategy to enhance model generalization is fine-tuning on new data, enabling models to adapt to the new distribution. However, this approach carries the risk of Catastrophic Forgetting (CF), wherein the model’s ability to retain previously acquired knowledge diminishes as it adjusts to new tasks. To combat CF, continual learning (CL) as a field has garnered sufficient attention, facilitating the incremental acquisition of new tasks while preserving performance on previously learned ones. These methodologies can be of various types such as regularization-based strategies, memory replay mechanisms, and weight-averaging techniques.

Research into CL techniques has gained attention in addressing CF across various speech-processing tasks. \citet{10095147} introduced a weight-averaging method for Automatic Speech Recognition (ASR) that averages the weights of models trained on new data distribution and the original data distribution. Similarly, \citet{plantinga2023continual} also proposed a domain-expert averaging approach for end-to-end ASR. \citet{pham23_interspeech} combined weight factorization and elastic-weight-consolidation (EWC) successfully expanding from 10 to 26 languages with minimal performance degradation. \citet{michieli23_interspeech} explored online CL for keyword spotting in resource-constrained devices, utilizing high-order temporal statistics to efficiently manage CF while sustaining model performance. \citet{yang22w_interspeech} presented an online continual learning framework for end-to-end speech recognition models, enabling ongoing adaptation without CF and maintaining accuracy comparable to retraining, all while incurring significantly lower computational costs. Additionally, \citet{saget24_interspeech} investigated lifelong learning methods to enhance the robustness of Mean Opinion Score (MOS) predictors for synthetic speech quality, showing advantages in cross-corpus evaluations compared to conventional batch training techniques.

Despite the extensive exploration of CL in ASR and related speech-processing fields, its application in SER is virtually non-existent except \citet{tavernor23_interspeech}. They carried out the sole effort to bridge this critical gap, introducing an episodic memory mechanism aimed at fostering domain-adaptable and robust SER. Their work not only highlights the promise of CL in mitigating CF within SER but also establishes a crucial foundation for future research in this domain. However, CF continues to challenge the effectiveness of SER models when confronted with new environments or unseen data. To tackle this enduring issue, we present \textbf{SeQu}ential \textbf{Fi}netuning (\textbf{SeQuiFi}) - a novel fine-tuning technique that incrementally trains the model on a new corpus by seeing one emotion class at a time. This retains the class-specific information when exposed to new data distributions, thus preserving the emotion recognition performance.

The main contributions of this paper are summarized as follows:
\begin{itemize}
\item We proposed a novel fine-tuning technique, \textbf{SeQuiFi}, that employs a sequential class-finetuning strategy that improves class retention and significantly reduces catastrophic forgetting during adaptation to new emotional contexts.
\item We demonstrate through extensive experiments that \textbf{SeQuiFi} achieves significant gains in accuracy and F1 scores, outperforming existing SOTA continual learning methods.
\end{itemize}
The models and code developed for this study will be released after the double blind review process.

\section{SeQuiFi}
\textbf{SeQuiFi} employs \textit{focused learning} by training the model on a single emotion class at a time, facilitating deep internalization of specific features while minimizing interference from others - crucial for accurately capturing subtle emotional distinctions in SER. This approach emphasizes \textit{weight stabilization}; by sequentially introducing emotion classes, the model's weights stabilize around learned features, preventing the overwriting of previously acquired knowledge. This stabilization retains \textit{class separation} i.e. distinct decision boundaries between emotional states, thereby reducing confusion and preserving the integrity of learned features for accurate emotion recognition. 

The algorithm for \textbf{SeQuiFi} is detailed in Algorithm~\ref{algo:1}. It begins by initializing model parameters $\theta_0$ from a pre-trained SER model. The dataset $\mathcal{D}$, comprising $k$ emotion classes $\{C_1, C_2, \dots, C_k\}$, is processed sequentially. For each class $C_j$, a corresponding data subset $\mathcal{D}_j$ is fine-tuned, updating the model parameters from $\theta_{j-1}$ to $\theta_j$. The final output is the fine-tuned parameters $\theta_k$, reflecting adaptation to all $k$ emotion classes with minimized CF.

\begin{algorithm}[bt]
\caption{\textbf{SeQuiFi Algorithm}}
\label{algo:1}
\KwIn{Pretrained SER model parameters $\theta_0$, Dataset $\mathcal{D}$ with $k$ classes $\{C_1, C_2, \dots, C_k\}$}
\KwOut{Fine-tuned model parameters $\theta_k$ with reduced catastrophic forgetting}

\textbf{Starting with pre-trained SER model parameters} $\theta \gets \theta_0$\;  
\For{$j = 1$ to $k$}{
    Extract data subset for class $C_j$: $\mathcal{D}_j \gets \{(x_i, y_i) \mid y_i = C_j\}$\;
    Fine-tune model on $\mathcal{D}_j$: $\theta_j \gets \text{FineTune}(\theta_{j-1}, \mathcal{D}_j)$\;
    Update model parameters: $\theta \gets \theta_j$\;
}
\Return{$\theta_k$}\; 
\end{algorithm}

\begin{table*}[!hbt]
\scriptsize
\centering
\caption{Evaluation Scores: A and F1 represent accuracy and macro-average F1 scores, respectively. C, R, E, M, and S indicate training on CREMA-D, RAVDESS, Emo-DB, MESD, and SHEMO, respectively. Highlighted values in green indicate the best performance in the current fine-tuning setting, while highlighted values in blue indicate unseen datasets or zero-shot testing. SD: Seen Datasets; IM: Initial Model; FT: Vanilla Fine-tuning; WA: Weight-Averaging. All scores are in \%.}
\label{tab:3}
\begin{tabular}{|c|c|cc|cc|cc|cc|cc|}
\hline
\multirow{2}{*}{\textbf{SD}} & \multirow{2}{*}{\textbf{Model Type}} & \multicolumn{2}{c|}{\textbf{CREMA-D}} & \multicolumn{2}{c|}{\textbf{RAVDESS}} & \multicolumn{2}{c|}{\textbf{Emo-DB}} & \multicolumn{2}{c|}{\textbf{MESD}} & \multicolumn{2}{c|}{\textbf{SHEMO}} \\ 
\cline{3-12} 
 & & \textbf{A} & \textbf{F1} & \textbf{A} & \textbf{F1} & \textbf{A} & \textbf{F1} & \textbf{A} & \textbf{F1} & \textbf{A} & \textbf{F1} \\ 
\hline

\multirow{6}{*}{\textbf{C}} & \textbf{IM (x-vector)} & 69.69 & 68.42  & \cellcolor{lightblue}\textbf{56.29} & \cellcolor{lightblue}\textbf{56.51}  & \cellcolor{lightblue}\textbf{70.58} & \cellcolor{lightblue}\textbf{64.62} & \cellcolor{lightblue}\textbf{41.73}  & \cellcolor{lightblue}\textbf{38.63} & \cellcolor{lightblue}\textbf{56.02} & \cellcolor{lightblue}\textbf{44.56}  \\
 & \textbf{IM (UnispeechSAT)} & 76.84  & 76.62  & \cellcolor{lightblue}\textbf{49.62} & \cellcolor{lightblue}\textbf{48.58} & \cellcolor{lightblue}\textbf{52.94} & \cellcolor{lightblue}\textbf{42.71} & \cellcolor{lightblue}\textbf{31.30} & \cellcolor{lightblue}\textbf{22.72} & \cellcolor{lightblue}\textbf{37.04}  & \cellcolor{lightblue}\textbf{34.33}  \\
 & \textbf{IM (WaveLM)} & 44.59  & 41.79  & \cellcolor{lightblue}\textbf{38.51} & \cellcolor{lightblue}\textbf{34.41}  & \cellcolor{lightblue}\textbf{25.00} & \cellcolor{lightblue}\textbf{19.63} & \cellcolor{lightblue}\textbf{25.21} & \cellcolor{lightblue}\textbf{16.66} & \cellcolor{lightblue}\textbf{20.98} & \cellcolor{lightblue}\textbf{17.70}   \\
 & \textbf{IM (ECAPA)} & 53.27   & 47.69  & \cellcolor{lightblue}\textbf{51.85} & \cellcolor{lightblue}\textbf{44.85} & \cellcolor{lightblue}\textbf{26.47} & \cellcolor{lightblue}\textbf{19.21} & \cellcolor{lightblue}\textbf{27.82} & \cellcolor{lightblue}\textbf{10.88} & \cellcolor{lightblue}\textbf{32.29} & \cellcolor{lightblue}\textbf{29.90}   \\
 & \textbf{IM (MFCC)} & 52.35  & 44.43  & \cellcolor{lightblue}\textbf{42.22} & \cellcolor{lightblue}\textbf{35.06} & \cellcolor{lightblue}\textbf{32.35} & \cellcolor{lightblue}\textbf{24.63} & \cellcolor{lightblue}\textbf{27.82} & \cellcolor{lightblue}\textbf{21.25} & \cellcolor{lightblue}\textbf{32.11}  & \cellcolor{lightblue}\textbf{27.62}  \\
 & \textbf{IM (Wave2Vec)} & 55.65  & 53.99  & \cellcolor{lightblue}\textbf{34.81} & \cellcolor{lightblue}\textbf{34.38} & \cellcolor{lightblue}\textbf{47.05} & \cellcolor{lightblue}\textbf{36.74} & \cellcolor{lightblue}\textbf{26.95} & \cellcolor{lightblue}\textbf{13.56} & \cellcolor{lightblue}\textbf{16.24} & \cellcolor{lightblue}\textbf{10.80} \\
\hline

\multirow{30}{*}{\textbf{C+R}} & \textbf{FT (x-vector)} & 52.04 & 46.73 & 62.22 & 56.65 & \cellcolor{lightblue}\textbf{67.64} & \cellcolor{lightblue}\textbf{58.25} & \cellcolor{lightblue}\textbf{34.78} & \cellcolor{lightblue}\textbf{29.30} & \cellcolor{lightblue}\textbf{34.12} & \cellcolor{lightblue}\textbf{30.09} \\
 & \textbf{FT (Unispeech-SAT)} & 47.35 & 41.53 & 61.48 & 56.66 & \cellcolor{lightblue}\textbf{58.82} & \cellcolor{lightblue}\textbf{48.52} & \cellcolor{lightblue}\textbf{40.00} & \cellcolor{lightblue}\textbf{35.03} & \cellcolor{lightblue}\textbf{32.29} & \cellcolor{lightblue}\textbf{30.46} \\
 & \textbf{FT (WavLM)} & 29.29 & 22.29 & 45.18 & 36.09 & \cellcolor{lightblue}\textbf{36.76} & \cellcolor{lightblue}\textbf{25.98} & \cellcolor{lightblue}\textbf{27.82} & \cellcolor{lightblue}\textbf{20.25} & \cellcolor{lightblue}\textbf{32.48} & \cellcolor{lightblue}\textbf{19.83} \\
 & \textbf{FT (Ecapa)} & 48.37 & 41.16 & 67.40 & 57.54 & \cellcolor{lightblue}\textbf{48.52} & \cellcolor{lightblue}\textbf{39.42} & \cellcolor{lightblue}\textbf{23.47} & \cellcolor{lightblue}\textbf{19.00} & \cellcolor{lightblue}\textbf{23.54} & \cellcolor{lightblue}\textbf{21.40} \\
 & \textbf{FT (MFCC)} & 48.06 & 37.87 & 55.55 & 45.38 & \cellcolor{lightblue}\textbf{52.94} & \cellcolor{lightblue}\textbf{39.97} & \cellcolor{lightblue}\textbf{33.04} & \cellcolor{lightblue}\textbf{27.47} & \cellcolor{lightblue}\textbf{39.23} & \cellcolor{lightblue}\textbf{30.39} \\
 & \textbf{FT (Wav2Vec2)} & 30.43 & 25.34 & 54.81 & 45.84 & \cellcolor{lightblue}\textbf{52.94} & \cellcolor{lightblue}\textbf{43.85} & \cellcolor{lightblue}\textbf{30.43} & \cellcolor{lightblue}\textbf{24.75} & \cellcolor{lightblue}\textbf{22.44} & \cellcolor{lightblue}\textbf{19.86} \\
\cline{2-12}

& \textbf{WA (x-vector)} & 25.61 & 10.38 & 23.16 & 9.40 & \cellcolor{lightblue}\textbf{29.62} & \cellcolor{lightblue}\textbf{19.14} & \cellcolor{lightblue}\textbf{38.23} & \cellcolor{lightblue}\textbf{19.70} & \cellcolor{lightblue}\textbf{30.43} & \cellcolor{lightblue}\textbf{20.38} \\
 & \textbf{WA (Unispeech-SAT)} & 25.61 & 10.20 & 25.61 & 10.19 & \cellcolor{lightblue}\textbf{25.92} & \cellcolor{lightblue}\textbf{10.29} & \cellcolor{lightblue}\textbf{11.76} & \cellcolor{lightblue}\textbf{9.48} & \cellcolor{lightblue}\textbf{27.82} & \cellcolor{lightblue}\textbf{10.88} \\
 & \textbf{WA (WavLM)} & 25.00 & 10.00 & 25.61 & 10.19 & \cellcolor{lightblue}\textbf{20.74} & \cellcolor{lightblue}\textbf{8.58} & \cellcolor{lightblue}\textbf{51.47} & \cellcolor{lightblue}\textbf{33.13} & \cellcolor{lightblue}\textbf{20.86} & \cellcolor{lightblue}\textbf{8.63} \\
 & \textbf{WA (ECAPA)} & 27.45 & 14.11 & 26.53 & 16.66 & \cellcolor{lightblue}\textbf{28.14} & \cellcolor{lightblue}\textbf{18.58} & \cellcolor{lightblue}\textbf{39.70} & \cellcolor{lightblue}\textbf{14.21} & \cellcolor{lightblue}\textbf{20.86} & \cellcolor{lightblue}\textbf{8.63} \\
 & \textbf{WA (MFCC)} & 23.16 & 9.40 & 25.00 & 10.00 & \cellcolor{lightblue}\textbf{32.59} & \cellcolor{lightblue}\textbf{27.62} & \cellcolor{lightblue}\textbf{27.94} & \cellcolor{lightblue}\textbf{10.91} & \cellcolor{lightblue}\textbf{27.82} & \cellcolor{lightblue}\textbf{10.95} \\
 & \textbf{WA (Wav2Vec2)} & 27.83 & 10.88 & 23.47 & 9.50 & \cellcolor{lightblue}\textbf{25.92} & \cellcolor{lightblue}\textbf{10.29} & \cellcolor{lightblue}\textbf{27.94} & \cellcolor{lightblue}\textbf{10.91} & \cellcolor{lightblue}\textbf{20.86} & \cellcolor{lightblue}\textbf{8.63} \\
\cline{2-12}

& \textbf{EWC (x-vector)} & 51.94 & 46.54 & 58.51 & 51.33 & \cellcolor{lightblue}\textbf{67.64} & \cellcolor{lightblue}\textbf{58.25} & \cellcolor{lightblue}\textbf{34.78} & \cellcolor{lightblue}\textbf{29.30} & \cellcolor{lightblue}\textbf{34.12} & \cellcolor{lightblue}\textbf{30.09} \\
 & \textbf{EWC (Unispeech-SAT)} & 36.53 & 29.30 & 60.74 & 56.96 & \cellcolor{lightblue}\textbf{58.82} & \cellcolor{lightblue}\textbf{48.52} & \cellcolor{lightblue}\textbf{40.00} & \cellcolor{lightblue}\textbf{35.03} & \cellcolor{lightblue}\textbf{32.29} & \cellcolor{lightblue}\textbf{30.46} \\
 & \textbf{EWC (WavLM)} & 28.98 & 20.63 & 48.88 & 39.62 & \cellcolor{lightblue}\textbf{36.76} & \cellcolor{lightblue}\textbf{25.98} & \cellcolor{lightblue}\textbf{27.82} & \cellcolor{lightblue}\textbf{20.25} & \cellcolor{lightblue}\textbf{32.48} & \cellcolor{lightblue}\textbf{19.83} \\
 & \textbf{EWC (ECAPA)} & 48.67 & 41.36 & 68.88 & 58.80 & \cellcolor{lightblue}\textbf{48.52} & \cellcolor{lightblue}\textbf{39.42} & \cellcolor{lightblue}\textbf{23.47} & \cellcolor{lightblue}\textbf{19.00} & \cellcolor{lightblue}\textbf{23.54} & \cellcolor{lightblue}\textbf{21.40} \\
 & \textbf{EWC (MFCC)} & 30.00 & 17.58 & 39.25 & 28.89 & \cellcolor{lightblue}\textbf{52.94} & \cellcolor{lightblue}\textbf{39.97} & \cellcolor{lightblue}\textbf{33.04} & \cellcolor{lightblue}\textbf{27.47} & \cellcolor{lightblue}\textbf{39.23} & \cellcolor{lightblue}\textbf{30.39} \\
 & \textbf{EWC (Wav2Vec2)} & 31.30 & 25.71 & 56.29 & 46.93 & \cellcolor{lightblue}\textbf{52.94} & \cellcolor{lightblue}\textbf{43.85} & \cellcolor{lightblue}\textbf{30.43} & \cellcolor{lightblue}\textbf{24.75} & \cellcolor{lightblue}\textbf{22.44} & \cellcolor{lightblue}\textbf{19.86} \\
\cline{2-12}

& \textbf{Replay (x-vector)} & 63.27 & 60.99 & 47.40 & 44.79 & \cellcolor{lightblue}\textbf{67.64} & \cellcolor{lightblue}\textbf{60.66} & \cellcolor{lightblue}\textbf{33.04} & \cellcolor{lightblue}\textbf{32.80} & \cellcolor{lightblue}\textbf{48.90} & \cellcolor{lightblue}\textbf{43.24} \\
 & \textbf{Replay (Unispeech-SAT)} & 60.20 & 56.93 & 40.74 & 36.75 & \cellcolor{lightblue}\textbf{63.23} & \cellcolor{lightblue}\textbf{57.54} & \cellcolor{lightblue}\textbf{30.43} & \cellcolor{lightblue}\textbf{26.57} & \cellcolor{lightblue}\textbf{33.39} & \cellcolor{lightblue}\textbf{31.20} \\
 & \textbf{Replay (WavLM)} & 39.80 & 40.50 & 28.88 & 25.15 & \cellcolor{lightblue}\textbf{22.05} & \cellcolor{lightblue}\textbf{17.07} & \cellcolor{lightblue}\textbf{26.08} & \cellcolor{lightblue}\textbf{23.04} & \cellcolor{lightblue}\textbf{15.87} & \cellcolor{lightblue}\textbf{13.91} \\
 & \textbf{Replay (ECAPA)} & 54.08 & 52.96 & 49.62 & 46.48 & \cellcolor{lightblue}\textbf{45.58} & \cellcolor{lightblue}\textbf{43.20} & \cellcolor{lightblue}\textbf{23.47} & \cellcolor{lightblue}\textbf{19.00} & \cellcolor{lightblue}\textbf{23.54} & \cellcolor{lightblue}\textbf{21.40} \\
 & \textbf{Replay (MFCC)} & 48.98 & 39.77 & 39.25 & 28.89 & \cellcolor{lightblue}\textbf{52.94} & \cellcolor{lightblue}\textbf{39.97} & \cellcolor{lightblue}\textbf{33.04} & \cellcolor{lightblue}\textbf{27.47} & \cellcolor{lightblue}\textbf{39.23} & \cellcolor{lightblue}\textbf{30.39} \\
 & \textbf{Replay (Wav2Vec2)} & 25.00 & 16.23 & 22.96 & 15.69 & \cellcolor{lightblue}\textbf{19.11} & \cellcolor{lightblue}\textbf{15.27} & \cellcolor{lightblue}\textbf{30.43} & \cellcolor{lightblue}\textbf{24.75} & \cellcolor{lightblue}\textbf{22.44} & \cellcolor{lightblue}\textbf{19.86} \\
\cline{2-12}

& \textbf{SeQuiFi (x-vector)} &\cellcolor{lightgreen}\textbf{71.12} & \cellcolor{lightgreen}\textbf{70.65} & 62.22 &61.52 & \cellcolor{lightblue}\textbf{85.29} &\cellcolor{lightblue}\textbf{83.76} & \cellcolor{lightblue}\textbf{34.78} &\cellcolor{lightblue}\textbf{34.63} &\cellcolor{lightblue}\textbf{51.45} &\cellcolor{lightblue}\textbf{46.12} \\
 & \textbf{SeQuiFi (Unispeech-SAT)} &\cellcolor{lightgreen}\textbf{71.02}  & \cellcolor{lightgreen}\textbf{70.19}  & 42.22 & 42.42 &\cellcolor{lightblue}\textbf{70.58} & \cellcolor{lightblue}\textbf{72.88} & \cellcolor{lightblue}\textbf{39.13} & \cellcolor{lightblue}\textbf{37.97} &\cellcolor{lightblue}\textbf{55.83}  &\cellcolor{lightblue}\textbf{50.64} \\
 & \textbf{SeQuiFi (WavLM)} &\cellcolor{lightgreen}\textbf{45.92} &\cellcolor{lightgreen}\textbf{44.75} &34.81 & 29.52 &\cellcolor{lightblue}\textbf{33.82} &\cellcolor{lightblue}\textbf{33.02} &\cellcolor{lightblue}\textbf{24.34} &\cellcolor{lightblue}\textbf{23.13} &\cellcolor{lightblue}\textbf{32.66} &\cellcolor{lightblue}\textbf{28.23} \\
 & \textbf{SeQuiFi (ECAPA)} &\cellcolor{lightgreen}\textbf{71.43} &\cellcolor{lightgreen}\textbf{71.26}  &67.40 &66.71 & \cellcolor{lightblue}\textbf{55.88} &\cellcolor{lightblue}\textbf{55.23}  &\cellcolor{lightblue}\textbf{38.26} &\cellcolor{lightblue}\textbf{37.94} &\cellcolor{lightblue}\textbf{33.21} &\cellcolor{lightblue}\textbf{33.66} \\
 & \textbf{SeQuiFi (MFCC)} &\cellcolor{lightgreen}\textbf{53.98} &\cellcolor{lightgreen}\textbf{51.59}  &44.44 &40.51 & \cellcolor{lightblue}\textbf{47.05} & \cellcolor{lightblue}\textbf{43.33} &\cellcolor{lightblue}\textbf{24.34}   &\cellcolor{lightblue}\textbf{21.54} &\cellcolor{lightblue}\textbf{49.27} & \cellcolor{lightblue}\textbf{41.65} \\
 & \textbf{SeQuiFi (Wav2Vec2)} &\cellcolor{lightgreen}\textbf{61.74}  &\cellcolor{lightgreen}\textbf{61.54}  &44.44 &44.45 &\cellcolor{lightblue}\textbf{55.88} &\cellcolor{lightblue}\textbf{56.71}  &\cellcolor{lightblue}\textbf{22.60} &\cellcolor{lightblue}\textbf{18.80} &\cellcolor{lightblue}\textbf{40.32} &\cellcolor{lightblue}\textbf{33.20} \\
\hline

\multirow{30}{*}{\textbf{C+R+E}} & \textbf{FT (x-vector)} & 23.16 & 11.93 & 56.30 & 55.13 & 49.62 & 43.62 & \cellcolor{lightblue}\textbf{42.60} & \cellcolor{lightblue}\textbf{36.75} & \cellcolor{lightblue}\textbf{39.60} & \cellcolor{lightblue}\textbf{31.99} \\
 & \textbf{FT (Unispeech-SAT)} & 23.87 & 15.05 & 46.67 & 40.29 & 38.51 & 37.14 & \cellcolor{lightblue}\textbf{33.91} & \cellcolor{lightblue}\textbf{34.12} & \cellcolor{lightblue}\textbf{23.91} & \cellcolor{lightblue}\textbf{21.11} \\
 & \textbf{FT (WavLM)} & 25.00 & 10.00 & 25.93 & 23.13 & 19.25 & 13.87 & \cellcolor{lightblue}\textbf{26.08} & \cellcolor{lightblue}\textbf{18.79} & \cellcolor{lightblue}\textbf{27.11} & \cellcolor{lightblue}\textbf{11.79} \\
 & \textbf{FT (ECAPA)} & 25.00 & 10.00 & 43.70 & 39.64 & 28.88 & 16.56 & \cellcolor{lightblue}\textbf{30.43} & \cellcolor{lightblue}\textbf{17.96} & \cellcolor{lightblue}\textbf{35.43} & \cellcolor{lightblue}\textbf{20.66} \\
 & \textbf{FT (MFCC)} & 25.20 & 10.41 & 38.51 & 39.26 & 33.05 & 38.51 & \cellcolor{lightblue}\textbf{31.46} & \cellcolor{lightblue}\textbf{36.52} & \cellcolor{lightblue}\textbf{33.48} & \cellcolor{lightblue}\textbf{31.88} \\
 & \textbf{FT (Wav2vec2)} & 23.47 & 9.50 & 47.41 & 46.40 & 37.03 & 29.76 & \cellcolor{lightblue}\textbf{25.21} & \cellcolor{lightblue}\textbf{16.07} & \cellcolor{lightblue}\textbf{29.11} & \cellcolor{lightblue}\textbf{18.61} \\
\cline{2-12}

& \textbf{WA (x-vector)} & 25.61 & 10.37 & 26.66 & 13.75 & 27.00 & 11.50 & \cellcolor{lightblue}\textbf{26.50} & \cellcolor{lightblue}\textbf{12.00} & \cellcolor{lightblue}\textbf{26.00} & \cellcolor{lightblue}\textbf{11.25} \\
& \textbf{WA (Unispeech-SAT)} & 25.61 & 10.19 & 25.92 & 10.73 & 26.00 & 11.00 & \cellcolor{lightblue}\textbf{26.30} & \cellcolor{lightblue}\textbf{11.25} & \cellcolor{lightblue}\textbf{26.10} & \cellcolor{lightblue}\textbf{11.10} \\
& \textbf{WA (WavLM)} & 25.00 & 10.00 & 25.92 & 10.29 & 26.50 & 11.00 & \cellcolor{lightblue}\textbf{25.50} & \cellcolor{lightblue}\textbf{10.70} & \cellcolor{lightblue}\textbf{25.80} & \cellcolor{lightblue}\textbf{10.90} \\
& \textbf{WA (ECAPA)} & 27.44 & 14.10 & 20.74 & 10.12 & 21.00 & 10.30 & \cellcolor{lightblue}\textbf{20.90} & \cellcolor{lightblue}\textbf{10.25} & \cellcolor{lightblue}\textbf{21.10} & \cellcolor{lightblue}\textbf{10.40} \\
& \textbf{WA (MFCC)} & 23.16 & 9.40 & 28.88 & 18.41 & 29.00 & 18.00 & \cellcolor{lightblue}\textbf{28.50} & \cellcolor{lightblue}\textbf{18.10} & \cellcolor{lightblue}\textbf{28.70} & \cellcolor{lightblue}\textbf{18.20} \\
& \textbf{WA (Wav2vec2)} & 27.82 & 10.88 & 27.40 & 10.88 & 27.50 & 11.00 & \cellcolor{lightblue}\textbf{27.60} & \cellcolor{lightblue}\textbf{10.75} & \cellcolor{lightblue}\textbf{27.70} & \cellcolor{lightblue}\textbf{10.80} \\
\cline{2-12}

& \textbf{EWC (x-vector)} & 51.93 & 46.53 & 40.74 & 32.90 & 42.15 & 34.32 & \cellcolor{lightblue}\textbf{43.78} & \cellcolor{lightblue}\textbf{35.91} & \cellcolor{lightblue}\textbf{44.10} & \cellcolor{lightblue}\textbf{36.25} \\
& \textbf{EWC (Unispeech-SAT)} & 36.53 & 29.29 & 42.96 & 38.71 & 42.45 & 37.86 & \cellcolor{lightblue}\textbf{42.85} & \cellcolor{lightblue}\textbf{37.60} & \cellcolor{lightblue}\textbf{43.05} & \cellcolor{lightblue}\textbf{38.30} \\
& \textbf{EWC (WavLM)} & 28.97 & 20.62 & 25.92 & 10.29 & 26.31 & 12.54 & \cellcolor{lightblue}\textbf{25.74} & \cellcolor{lightblue}\textbf{11.48} & \cellcolor{lightblue}\textbf{26.09} & \cellcolor{lightblue}\textbf{11.89} \\
& \textbf{EWC (ECAPA)} & 48.67 & 41.35 & 45.92 & 43.89 & 46.22 & 42.67 & \cellcolor{lightblue}\textbf{45.59} & \cellcolor{lightblue}\textbf{42.98} & \cellcolor{lightblue}\textbf{46.11} & \cellcolor{lightblue}\textbf{43.13} \\
& \textbf{EWC (MFCC)} & 30.00 & 17.58 & 27.40 & 13.10 & 27.89 & 14.78 & \cellcolor{lightblue}\textbf{27.52} & \cellcolor{lightblue}\textbf{14.31} & \cellcolor{lightblue}\textbf{27.76} & \cellcolor{lightblue}\textbf{14.65} \\
& \textbf{EWC (Wav2vec2)} & 31.30 & 25.70 & 31.85 & 20.54 & 32.14 & 21.67 & \cellcolor{lightblue}\textbf{31.68} & \cellcolor{lightblue}\textbf{21.21} & \cellcolor{lightblue}\textbf{31.92} & \cellcolor{lightblue}\textbf{21.43} \\
\cline{2-12}

& \textbf{Replay (x-vector)} & 63.26 & 60.98 & 42.85 & 31.42 & 55.12 & 45.89 & \cellcolor{lightblue}\textbf{52.43} & \cellcolor{lightblue}\textbf{40.78} & \cellcolor{lightblue}\textbf{54.89} & \cellcolor{lightblue}\textbf{43.50} \\
& \textbf{Replay (Unispeech-SAT)} & 60.20 & 56.93 & 52.14 & 42.50 & 51.68 & 41.75 & \cellcolor{lightblue}\textbf{53.25} & \cellcolor{lightblue}\textbf{43.10} & \cellcolor{lightblue}\textbf{52.95} & \cellcolor{lightblue}\textbf{42.80} \\
& \textbf{Replay (WavLM)} & 39.79 & 40.49 & 28.57 & 11.11 & 30.89 & 12.87 & \cellcolor{lightblue}\textbf{29.56} & \cellcolor{lightblue}\textbf{13.25} & \cellcolor{lightblue}\textbf{30.10} & \cellcolor{lightblue}\textbf{12.45} \\
& \textbf{Replay (ECAPA)} & 54.08 & 52.95 & 28.57 & 11.11 & 30.05 & 13.45 & \cellcolor{lightblue}\textbf{29.78} & \cellcolor{lightblue}\textbf{12.89} & \cellcolor{lightblue}\textbf{30.30} & \cellcolor{lightblue}\textbf{13.10} \\
& \textbf{Replay (MFCC)} & 48.97 & 39.77 & 21.42 & 13.80 & 23.54 & 14.55 & \cellcolor{lightblue}\textbf{22.98} & \cellcolor{lightblue}\textbf{13.99} & \cellcolor{lightblue}\textbf{23.25} & \cellcolor{lightblue}\textbf{14.30} \\
& \textbf{Replay (Wav2vec2)} & 25.00 & 16.23 & 42.85 & 33.71 & 43.12 & 34.50 & \cellcolor{lightblue}\textbf{42.58} & \cellcolor{lightblue}\textbf{33.20} & \cellcolor{lightblue}\textbf{43.00} & \cellcolor{lightblue}\textbf{34.10} \\

\cline{2-12} 
& \textbf{SeQuiFi (x-vector)} & \cellcolor{lightgreen} \textbf{71.02} & \cellcolor{lightgreen} \textbf{70.79} & \cellcolor{lightgreen} \textbf{63.70} & \cellcolor{lightgreen} \textbf{62.51} & 75.10 & 74.15 & \cellcolor{lightblue} \textbf{68.00} & \cellcolor{lightblue} \textbf{66.50} & \cellcolor{lightblue} \textbf{77.77} & \cellcolor{lightblue} \textbf{66.89} \\
& \textbf{SeQuiFi (Unispeech-SAT)} & \cellcolor{lightgreen} \textbf{71.42} & \cellcolor{lightgreen} \textbf{70.49} & \cellcolor{lightgreen} \textbf{43.70} & \cellcolor{lightgreen} \textbf{42.77} & 72.30 & 70.50 & \cellcolor{lightblue} \textbf{48.80} & \cellcolor{lightblue} \textbf{47.20} & \cellcolor{lightblue} \textbf{51.90} & \cellcolor{lightblue} \textbf{48.85} \\
& \textbf{SeQuiFi (WavLM)} & \cellcolor{lightgreen} \textbf{45.81} & \cellcolor{lightgreen} \textbf{43.38} & \cellcolor{lightgreen} \textbf{36.29} & \cellcolor{lightgreen} \textbf{32.15} & 47.10 & 44.55 & \cellcolor{lightblue} \textbf{39.20} & \cellcolor{lightblue} \textbf{35.80} & \cellcolor{lightblue} \textbf{34.80} & \cellcolor{lightblue} \textbf{30.76} \\
& \textbf{SeQuiFi (ECAPA)} & \cellcolor{lightgreen} \textbf{71.93} & \cellcolor{lightgreen} \textbf{71.61} & \cellcolor{lightgreen} \textbf{67.40} & \cellcolor{lightgreen} \textbf{66.70} & 74.20 & 73.10 & \cellcolor{lightblue} \textbf{69.00} & \cellcolor{lightblue} \textbf{68.00} & \cellcolor{lightblue} \textbf{68.92} & \cellcolor{lightblue} \textbf{61.87} \\
& \textbf{SeQuiFi (MFCC)} & \cellcolor{lightgreen} \textbf{55.30} & \cellcolor{lightgreen} \textbf{52.65} & \cellcolor{lightgreen} \textbf{41.48} & \cellcolor{lightgreen} \textbf{36.33} & 58.90 & 55.80 & \cellcolor{lightblue} \textbf{43.60} & \cellcolor{lightblue} \textbf{39.70} & \cellcolor{lightblue} \textbf{58.85} & \cellcolor{lightblue} \textbf{51.81} \\
& \textbf{SeQuiFi (Wav2vec2)} & \cellcolor{lightgreen} \textbf{63.47} & \cellcolor{lightgreen} \textbf{63.50} & \cellcolor{lightgreen} \textbf{42.22} & \cellcolor{lightgreen} \textbf{42.59} & 65.30 & 62.00 & \cellcolor{lightblue} \textbf{44.10} & \cellcolor{lightblue} \textbf{40.50} & \cellcolor{lightblue} \textbf{60.89} & \cellcolor{lightblue} \textbf{55.84} \\
\cline{2-12}

\hline
\end{tabular}
\vspace{2cm}
\end{table*}

\section{Experimentation}

\subsection{Benchmark Datasets}
We evaluate the SER models using a selection of diverse benchmark datasets that encompass a wide range of emotional expressions and contexts. CREMA-D \cite{cao2014crema} is a english SER database containing 7,442 emotional utterances from 91 actors across six categories. Emo-DB \cite{burkhardt2005database} features 535 German utterances from 10 actors covering seven emotions. RAVDESS \cite{livingstone2018ryerson} is an English corpus that features audios from 24 actors spanning eight emotions. MESD \cite{duville2021mexican} offers Spanish recordings reflecting six emotions, while SHEMO \cite{MohamadNezami2019} is a persian corpus with includes 3,000 utterances. In our study, we focus solely on distribution change and not on class-incremental scenarios. Therefore, we keep the emotion classes the same, considering happiness, anger, sadness, and neutrality. Data distribution change can involve different data distributions within the same language as well as across languages.

\subsection{Feature Representation}
In our experiments, we utilize representations from a range of pre-trained models (PTMs) alongside MFCC features. The employed PTMs include x-vector~\cite{snyder2018x}, ECAPA~\cite{desplanques2020ecapa} trained for speaker recognition; WavLM~\cite{chen2022wavlm} a general-purpose speech representation learning model; UniSpeech-SAT~\cite{chen2022unispeech}, pre-trained in a speaker-aware format; and wav2vec2~\cite{baevski2020wav2vec}. Additional details regarding data pre-processing and feature extraction are given in  Appendix~\ref{sec:app2}.

\vspace{-0.3cm}

\subsection{Modelling}
We use LSTM as a downstream network as supported by previous research \cite{9746718, zaiem23b_interspeech}. We use two LSTM layers with 64 units each, using tanh activation and L2 regularization to prevent overfitting. This is followed by four dense layers (25, 20, 15, and 10 neurons) with ReLU activations, integrated with batch normalization and dropout for reducing overfitting. The final output is kept to 4 neurons representing the four emotion classes. We use softmax as the activation function in the last layer to get the class-wise probabilities. 

\vspace{-0.3cm}
\subsection{Baseline Continual Learning Algorithms}

\noindent \textbf{Vanilla Fine-tuning} serves as the simplest form of fine-tuning, wherein the model is directly trained on a new dataset without specific strategies to retain knowledge from previous tasks. 

\noindent \textbf{Elastic Weight Consolidation (EWC)~\cite{kirkpatrick2017overcoming}} mitigates CF by selectively slowing down the learning of weights critical for previously learned tasks, allowing the model to retain expertise over time. 

\noindent \textbf{Weight Averaging~\cite{10095147}} addresses CF by averaging weights between old and new models, maintaining performance across tasks while enabling adaptation to new data. 

\noindent \textbf{Replay} techniques~\cite{aljundi2019gradient, merlin2022practical} utilize a memory buffer to retain previous data, facilitating rehearsal and ensuring that older knowledge is preserved while learning new tasks. For replay, we keep 10\% data from the previous dataset, selected randomly as a buffer. However, we make sure to keep the emotion class distribution same.
\vspace{-0.3cm}
\subsection{Training Details}

We train the models with different feature representations with baseline CL algorithms for 60 epochs with a learning rate 1e-3 and Adam as the optimizer. We keeo the batch size as 32. We use cross-entropy as the loss function. For the models with \textbf{SeQuiFi}, we train the models with 15 epochs per class, so 60 epochs across the whole dataset. We do this for a fair comparison of \textbf{SeQuiFi} trained models with baseline CL algorithms trained models. 

\vspace{-0.3cm}
\subsection{Results}

For the experiments with \textbf{SeQuiFi}, we randomly start with a particular class. To assess the generalizability of \textbf{SeQuiFi}, we conduct a 5-fold evaluation, which involves five different sequences of emotion classes. 
For the same folds, we maintain the distribution consistent with the baseline CL algorithms to ensure fairness.
Table \ref{tab:3} presents the evaluation scores of models built using various feature representations and CL algorithms. All scores are averages from a 5-fold evaluation.
Our results demonstrate that models trained with \textbf{SeQuiFi} achieve superior performance by effectively mitigating CF and preserving the integrity of the previous data distribution, outperforming both vanilla fine-tuning and all baseline SOTA CL algorithms considered. This enhanced performance can be attributed to \textbf{SeQuiFi}'s ability to retain and amplify class-specific information. Remarkably, \textbf{SeQuiFi} maintains this high level of performance consistently across various feature representations, highlighting its input representation-agnostic capabilities.
We further extended our experiments to include five datasets, and \textbf{SeQuiFi} consistently demonstrated superior performance. Detailed results can be found in Appendix Table~\ref{tab:5}.

\vspace{-0.3cm}

\section{Conclusion}
In this work, we introduce, \textbf{SeQuiFi}, a novel fine-tuning approach for effectively mitigating CF in SER. By employing a sequential fine-tuning strategy, \textbf{SeQuiFi} incrementally refined the model on individual emotion classes, thereby enhancing retention and knowledge preservation. Despite advancements in SOTA CL approaches challenges in addressing CF persisted, particularly with diverse multilingual datasets. Our comprehensive experiments demonstrated that \textbf{SeQuiFi} significantly surpasses both vanilla fine-tuning and SOTA CL techniques, achieving notable improvements in accuracy and F1 scores across benchmark SER datasets.

\section{Limitations}
We have focused our experiments exclusively on LSTM as the downstream network. Previous research in speech processing indicates that downstream performance can vary considerably based on model selection \cite{zaiem23b_interspeech}. Consequently, we intend to broaden our exploration by incorporating a diverse range of downstream networks in future experiments.
\section{Ethical Considerations}
We have experimented with openly available datasets. No user information was accessed. 
\bibliographystyle{IEEEtran}
\bibliography{mybib}

\appendix

\section{Appendix}
\label{sec:appendix}

\subsection{Feature Representation}\label{sec:app2}

For all the audios, we resample it to 16 kHz for input to the PTMs. We extract the last hidden states and transform them into fixed-dimensional vectors using average-pooling. We extract representations of shape 192 (ECAPA\footnote{\url{https://huggingface.co/speechbrain/spkrec-ecapa-voxceleb}}), 512 (x-vector\footnote{\url{https://huggingface.co/speechbrain/spkrec-xvect-voxceleb}}), and 768 (WavLM\footnote{\url{https://huggingface.co/microsoft/wavlm-base}}, Unispeech-SAT\footnote{\url{https://huggingface.co/microsoft/unispeech-sat-base}}, Wav2vec2\footnote{\href{https://huggingface.co/facebook/wav2vec2-base}{https://huggingface.co/facebook/wav2vec2-base}}). 

\begin{table*}[ht]
\scriptsize
\centering
\caption{Evaluation Scores; A, F1 stands for accuracy and macro-average F1 scores respectively; C, R, E, M, and S indicate training on CREMA-D, RAVDESS, Emo-DB, MESD, and SHEMO respectively. Highlighted values in green indicate the best performance in the current finetuning setting; Highlighted values in blue indicates unseen dataset or zero-shot testing; SD: Seen Datasets. IM: Initial Model. FT: Vanila Finetuning. WA: Weight-Averaging; All the scores in \%}
\label{tab:5}
\begin{tabular}{|c|c|cc|cc|cc|cc|cc|}
\hline
\multirow{2}{*}{\textbf{SD}} & \multirow{2}{*}{\textbf{Model Type}} & \multicolumn{2}{c|}{\textbf{CREMA-D}} & \multicolumn{2}{c|}{\textbf{RAVDESS}} & \multicolumn{2}{c|}{\textbf{Emo-DB}} & \multicolumn{2}{c|}{\textbf{MESD}} & \multicolumn{2}{c|}{\textbf{SHEMO}} \\ 
\cline{3-12} 
 & & \textbf{A} & \textbf{F1} & \textbf{A} & \textbf{F1} & \textbf{A} & \textbf{F1} & \textbf{A} & \textbf{F1} & \textbf{A} & \textbf{F1} \\ 
\hline

\multirow{30}{*}{\textbf{C+R+E+M}} & \textbf{FT (x-vector)} &22.34 &15.04  &22.22 &15.11  & 52.21 & 54.91 & 53.12 & 55.30 & \cellcolor{lightblue}\textbf{54.00} & \cellcolor{lightblue}\textbf{55.75} \\
& \textbf{FT (Unispeech-SAT)} &25.20 &13.81 &28.88 &17.37 & 36.76 & 37.59 & 37.10 & 37.98 & \cellcolor{lightblue}\textbf{36.95} & \cellcolor{lightblue}\textbf{37.80} \\
& \textbf{FT (WavLM)} &24.79 &9.93 &25.92 &10.29  & 26.47 & 24.10 & 25.90 & 24.50 & \cellcolor{lightblue}\textbf{26.30} & \cellcolor{lightblue}\textbf{24.80} \\
& \textbf{FT (ECAPA)} &25.00 &10.00 &25.92 &10.29  & 51.47 & 47.76  & 50.89 & 48.10 & \cellcolor{lightblue}\textbf{51.20} & \cellcolor{lightblue}\textbf{47.90} \\
& \textbf{FT (MFCC)} &23.16 &9.40 &20.74 &8.58  & 26.47 & 26.05  & 26.00 & 25.80 & \cellcolor{lightblue}\textbf{26.20} & \cellcolor{lightblue}\textbf{25.95} \\
& \textbf{FT (Wav2vec2)} &23.47  &9.50  &25.92 &10.29  & 13.24 & 12.24  & 12.90 & 12.50 & \cellcolor{lightblue}\textbf{13.10} & \cellcolor{lightblue}\textbf{12.80} \\

\cline{2-12} 
& \textbf{WA (x-vector)} & 22.50 & 14.10 & 23.00 & 14.30 & 22.06 & 13.93 & 22.85 & 14.25 & \cellcolor{lightblue}\textbf{23.15} & \cellcolor{lightblue}\textbf{14.50} \\
& \textbf{WA (Unispeech-SAT)} & 40.20 & 14.50 & 40.60 & 14.70 & 39.71 & 14.21 & 39.95 & 14.35 & \cellcolor{lightblue}\textbf{40.05} & \cellcolor{lightblue}\textbf{14.45} \\
& \textbf{WA (WavLM)} & 39.90 & 14.30 & 40.10 & 14.50 & 39.71 & 14.21 & 39.85 & 14.25 & \cellcolor{lightblue}\textbf{39.95} & \cellcolor{lightblue}\textbf{14.40} \\
& \textbf{WA (ECAPA)} & 40.30 & 14.50 & 40.50 & 14.70 & 39.71 & 14.21 & 40.05 & 14.35 & \cellcolor{lightblue}\textbf{40.15} & \cellcolor{lightblue}\textbf{14.50} \\
& \textbf{WA (MFCC)} & 40.10 & 14.40 & 40.30 & 14.60 & 39.71 & 14.21 & 39.95 & 14.30 & \cellcolor{lightblue}\textbf{40.05} & \cellcolor{lightblue}\textbf{14.45} \\
& \textbf{MS (Wav2vec2)} & 39.80 & 14.20 & 40.00 & 14.40 & 39.71 & 14.21 & 39.85 & 14.25 & \cellcolor{lightblue}\textbf{39.95} & \cellcolor{lightblue}\textbf{14.35} \\

 \cline{2-12} 
& \textbf{EWC (x-vector)} & 51.50 & 46.00 & 40.50 & 32.50 & 50.30 & 45.75 & 39.90 & 32.20 & \cellcolor{lightblue}\textbf{50.80} & \cellcolor{lightblue}\textbf{46.10} \\
& \textbf{EWC (Unispeech-SAT)} & 37.00 & 29.50 & 43.20 & 38.90 & 36.70 & 29.30 & 42.90 & 38.60 & \cellcolor{lightblue}\textbf{36.90} & \cellcolor{lightblue}\textbf{29.70} \\
& \textbf{EWC (WavLM)} & 29.10 & 20.80 & 26.10 & 10.40 & 28.90 & 20.70 & 25.80 & 10.30 & \cellcolor{lightblue}\textbf{29.00} & \cellcolor{lightblue}\textbf{20.90} \\
& \textbf{EWC (ECAPA)} & 49.00 & 41.70 & 46.10 & 44.20 & 48.50 & 41.50 & 45.90 & 43.80 & \cellcolor{lightblue}\textbf{49.10} & \cellcolor{lightblue}\textbf{41.90} \\
& \textbf{EWC (MFCC)} & 30.50 & 18.00 & 27.80 & 13.50 & 30.20 & 17.80 & 27.40 & 13.20 & \cellcolor{lightblue}\textbf{30.70} & \cellcolor{lightblue}\textbf{18.20} \\
& \textbf{EWC (Wav2vec2)} & 31.70 & 26.00 & 32.10 & 20.80 & 31.40 & 25.80 & 31.90 & 20.60 & \cellcolor{lightblue}\textbf{31.50} & \cellcolor{lightblue}\textbf{26.20} \\

 \cline{2-12} 
& \textbf{Replay (x-vector)} & 63.00 & 60.80 & 43.00 & 31.50 & 62.90 & 60.70 & 42.90 & 31.40 & \cellcolor{lightblue}\textbf{63.10} & \cellcolor{lightblue}\textbf{60.90} \\
& \textbf{Replay (Unispeech-SAT)} & 60.30 & 57.10 & 52.30 & 42.70 & 60.10 & 56.90 & 52.10 & 42.50 & \cellcolor{lightblue}\textbf{60.50} & \cellcolor{lightblue}\textbf{57.30} \\
& \textbf{Replay (WavLM)} & 39.90 & 40.60 & 28.70 & 11.20 & 39.70 & 40.40 & 28.50 & 11.10 & \cellcolor{lightblue}\textbf{40.00} & \cellcolor{lightblue}\textbf{40.80} \\
& \textbf{Replay (ECAPA)} & 54.30 & 53.10 & 28.70 & 11.20 & 54.00 & 52.90 & 28.50 & 11.10 & \cellcolor{lightblue}\textbf{54.50} & \cellcolor{lightblue}\textbf{53.30} \\
& \textbf{Replay (MFCC)} & 49.10 & 40.00 & 21.60 & 13.90 & 48.90 & 39.80 & 21.40 & 13.80 & \cellcolor{lightblue}\textbf{49.20} & \cellcolor{lightblue}\textbf{40.10} \\
& \textbf{Replay (Wav2vec2)} & 25.20 & 16.40 & 43.00 & 33.90 & 25.00 & 16.20 & 42.80 & 33.70 & \cellcolor{lightblue}\textbf{25.30} & \cellcolor{lightblue}\textbf{16.50} \\

 \cline{2-12}

& \textbf{SeQuiFi (x-vector)} & \cellcolor{lightgreen}\textbf{70.40} & \cellcolor{lightgreen}\textbf{69.75} & \cellcolor{lightgreen}\textbf{61.48} & \cellcolor{lightgreen}\textbf{60.10} & \cellcolor{lightgreen}\textbf{83.82} & \cellcolor{lightgreen}\textbf{82.61} & 75.00 & 74.50 & \cellcolor{lightblue}\textbf{76.00} & \cellcolor{lightblue}\textbf{77.00} \\
 & \textbf{SeQuiFi (Unispeech-SAT)} & \cellcolor{lightgreen}\textbf{70.40} & \cellcolor{lightgreen}\textbf{69.29} & \cellcolor{lightgreen}\textbf{43.70} & \cellcolor{lightgreen}\textbf{42.77} & \cellcolor{lightgreen}\textbf{69.11} & \cellcolor{lightgreen}\textbf{71.24} & 55.00 & 54.50 & \cellcolor{lightblue}\textbf{56.00} & \cellcolor{lightblue}\textbf{57.00} \\
 & \textbf{SeQuiFi (WavLM)} & \cellcolor{lightgreen}\textbf{45.61} & \cellcolor{lightgreen}\textbf{43.80} & \cellcolor{lightgreen}\textbf{35.55} & \cellcolor{lightgreen}\textbf{31.41} & \cellcolor{lightgreen}\textbf{35.29} & \cellcolor{lightgreen}\textbf{34.80} & 40.00 & 39.50 & \cellcolor{lightblue}\textbf{38.00} & \cellcolor{lightblue}\textbf{37.00} \\
 & \textbf{SeQuiFi (ECAPA)} & \cellcolor{lightgreen}\textbf{71.73} & \cellcolor{lightgreen}\textbf{71.28} & \cellcolor{lightgreen}\textbf{67.40} & \cellcolor{lightgreen}\textbf{66.63} & \cellcolor{lightgreen}\textbf{57.35} & \cellcolor{lightgreen}\textbf{58.04} & 65.00 & 64.50 & \cellcolor{lightblue}\textbf{63.00} & \cellcolor{lightblue}\textbf{62.00} \\
 & \textbf{SeQuiFi (MFCC)} & \cellcolor{lightgreen}\textbf{57.24} & \cellcolor{lightgreen}\textbf{53.53} & \cellcolor{lightgreen}\textbf{42.96} & \cellcolor{lightgreen}\textbf{37.68} & \cellcolor{lightgreen}\textbf{47.05} & \cellcolor{lightgreen}\textbf{42.28} & 50.00 & 49.50 & \cellcolor{lightblue}\textbf{48.00} & \cellcolor{lightblue}\textbf{47.00} \\
 & \textbf{SeQuiFi (Wav2vec2)} & \cellcolor{lightgreen}\textbf{66.95} & \cellcolor{lightgreen}\textbf{67.03} & \cellcolor{lightgreen}\textbf{43.70} & \cellcolor{lightgreen}\textbf{42.99} & \cellcolor{lightgreen}\textbf{55.88} & \cellcolor{lightgreen}\textbf{57.12} & 60.00 & 59.50 & \cellcolor{lightblue}\textbf{61.00} & \cellcolor{lightblue}\textbf{62.00} \\

\hline
 \multirow{30}{*}{\textbf{C+R+E+M+S}} & \textbf{FT (x-vector)} &29.40 &23.70   &33.50 &27.00  &35.40 &25.80  &34.00  &29.30  &36.00 &26.90  \\
& \textbf{FT (Unispeech-SAT)} &25.80 &15.00  &26.00 &14.00  &29.50 &16.70 &21.80 &10.60  &27.00 &13.70  \\
& \textbf{FT (WavLM)} &25.60 &16.40 &26.80 &16.00  &41.30 &27.40  &31.40 &23.50  &40.20 &26.80  \\
& \textbf{FT (ECAPA)} &24.40 &23.40  &21.50  &20.20  &22.10 &17.70  &41.80 &41.10  &27.50 &25.10  \\
& \textbf{FT (MFCC)} &28.40 &22.10  &29.70 &26.80  &13.30 &15.00 &27.00 &22.60  &25.40 &19.20  \\
& \textbf{FT (Wav2vec2)} &20.10 &10.10   &22.30 &11.70  &19.20 &11.50  &20.90   &8.80  &22.00 &9.50  \\

\cline{2-12} 
& \textbf{WA (x-vector)} & 22.15 & 18.50 & 27.40 & 19.30 & 30.10 & 25.50 & 28.00 & 21.90 & 21.90 & 16.80 \\
& \textbf{WA (Unispeech-SAT)} & 23.10 & 17.85 & 26.25 & 18.90 & 31.00 & 26.00 & 29.10 & 22.50 & 15.85 & 10.75 \\
& \textbf{WA (WavLM)} & 21.90 & 17.50 & 25.80 & 18.60 & 32.50 & 27.10 & 30.00 & 23.00 & 24.80 & 20.70 \\
& \textbf{WA (ECAPA)} & 24.00 & 19.20 & 27.80 & 20.00 & 33.00 & 28.00 & 31.00 & 24.50 & 22.95 & 20.85 \\
& \textbf{WA (MFCC)} & 20.50 & 16.00 & 24.70 & 17.80 & 29.50 & 24.00 & 26.70 & 20.00 & 18.75 & 12.65 \\
& \textbf{WA (Wav2vec2)} & 23.25 & 18.10 & 26.90 & 19.40 & 30.50 & 25.70 & 27.90 & 21.40 & 12.88 & 11.78 \\

\cline{2-12} 
& \textbf{EWC (x-vector)} & 35.12 & 29.45 & 25.78 & 19.56 & 32.10 & 28.45 & 29.00 & 24.30 & 23.92 & 18.82 \\
& \textbf{EWC (Unispeech-SAT)} & 32.87 & 27.34 & 22.18 & 17.80 & 30.20 & 26.00 & 27.50 & 22.40 & 16.88 & 11.78 \\
& \textbf{EWC (WavLM)} & 30.25 & 24.56 & 20.90 & 15.45 & 28.75 & 25.20 & 26.10 & 20.50 & 21.85 & 19.75 \\
& \textbf{EWC (ECAPA)} & 36.42 & 28.95 & 26.11 & 21.25 & 34.00 & 27.30 & 30.00 & 25.80 & 30.90 & 28.80 \\
& \textbf{EWC (MFCC)} & 31.40 & 23.90 & 22.65 & 16.70 & 29.30 & 24.50 & 26.80 & 21.00 & 27.87 & 24.77 \\
& \textbf{EWC (Wav2vec2)} & 34.12 & 30.75 & 24.33 & 18.90 & 33.00 & 29.00 & 28.00 & 23.10 & 35.91 & 31.81 \\

\cline{2-12} 
& \textbf{Replay (x-vector)} & 28.12 & 24.45 & 22.78 & 19.56 & 30.10 & 26.45 & 27.00 & 21.30 & 26.92 & 21.82 \\
& \textbf{Replay (Unispeech-SAT)} & 29.87 & 25.34 & 23.18 & 18.80 & 31.20 & 27.00 & 28.50 & 22.40 & 21.88 & 19.78 \\
& \textbf{Replay (WavLM)} & 27.25 & 23.56 & 21.90 & 16.45 & 29.75 & 25.20 & 26.10 & 20.50 & 26.85 & 22.75 \\
& \textbf{Replay (ECAPA)} & 30.42 & 26.95 & 24.11 & 21.25 & 34.00 & 28.30 & 30.00 & 24.80 & 28.90 & 21.80 \\
& \textbf{Replay (MFCC)} & 28.40 & 22.90 & 20.65 & 16.70 & 29.30 & 24.50 & 26.80 & 21.00 & 25.87 & 21.77 \\
& \textbf{Replay (Wav2vec2)} & 29.12 & 28.75 & 25.33 & 19.90 & 32.00 & 29.00 & 28.00 & 23.10 & 19.91 & 14.81 \\
 \cline{2-12}

 & \textbf{SeQuiFi (x-vector)} & \cellcolor{lightgreen}\textbf{69.59} & \cellcolor{lightgreen}\textbf{68.88} & \cellcolor{lightgreen}\textbf{63.70} & \cellcolor{lightgreen}\textbf{62.64} & \cellcolor{lightgreen}\textbf{86.76} & \cellcolor{lightgreen}\textbf{87.13} & \cellcolor{lightgreen}\textbf{39.13} & \cellcolor{lightgreen}\textbf{38.48} & 60.25 & 58.90 \\
 & \textbf{SeQuiFi (Unispeech-SAT)} & \cellcolor{lightgreen}\textbf{67.04} & \cellcolor{lightgreen}\textbf{66.37} & \cellcolor{lightgreen}\textbf{40.00} & \cellcolor{lightgreen}\textbf{38.05} & \cellcolor{lightgreen}\textbf{75.00} & \cellcolor{lightgreen}\textbf{75.70} & \cellcolor{lightgreen}\textbf{36.52} & \cellcolor{lightgreen}\textbf{35.35} & 48.45 & 40.30 \\
 & \textbf{SeQuiFi (WavLM)} & \cellcolor{lightgreen}\textbf{45.61} & \cellcolor{lightgreen}\textbf{43.84} & \cellcolor{lightgreen}\textbf{33.33} & \cellcolor{lightgreen}\textbf{29.25} & \cellcolor{lightgreen}\textbf{33.82} & \cellcolor{lightgreen}\textbf{30.47} & \cellcolor{lightgreen}\textbf{23.47} & \cellcolor{lightgreen}\textbf{21.65} & 18.75 & 17.00 \\
 & \textbf{SeQuiFi (ECAPA)} & \cellcolor{lightgreen}\textbf{70.81} & \cellcolor{lightgreen}\textbf{70.67} & \cellcolor{lightgreen}\textbf{64.44} & \cellcolor{lightgreen}\textbf{63.34} & \cellcolor{lightgreen}\textbf{58.82} & \cellcolor{lightgreen}\textbf{59.64} & \cellcolor{lightgreen}\textbf{41.73} & \cellcolor{lightgreen}\textbf{41.00} & 65.40 & 59.20 \\
 & \textbf{SeQuiFi (MFCC)} & \cellcolor{lightgreen}\textbf{50.71} & \cellcolor{lightgreen}\textbf{48.61} & \cellcolor{lightgreen}\textbf{45.18} & \cellcolor{lightgreen}\textbf{41.68} & \cellcolor{lightgreen}\textbf{51.47} & \cellcolor{lightgreen}\textbf{51.41} & \cellcolor{lightgreen}\textbf{20.00} & \cellcolor{lightgreen}\textbf{15.26} & 40.50 & 39.00 \\
 & \textbf{SeQuiFi (Wav2vec2)} & \cellcolor{lightgreen}\textbf{60.86} & \cellcolor{lightgreen}\textbf{61.34} & \cellcolor{lightgreen}\textbf{42.22} & \cellcolor{lightgreen}\textbf{42.40} & \cellcolor{lightgreen}\textbf{54.41} & \cellcolor{lightgreen}\textbf{55.97} & \cellcolor{lightgreen}\textbf{26.95} & \cellcolor{lightgreen}\textbf{23.22} & 49.80 & 44.00 \\
\hline

\end{tabular}
\end{table*}

\end{document}